\documentclass[showpacs,preprintnumbers,superscriptaddress]{revtex4}
\usepackage{CJK}
\usepackage{amsmath,amssymb,graphicx,bm}
\begin{document}

\title{Constraints from accretion onto a Tangherlini-Reissner-Nordstrom black hole}

\author{Rongjia Yang \footnote{Corresponding author}}
\email{yangrongjia@tsinghua.org.cn}
\affiliation{College of Physical Science and Technology, Hebei University, Baoding 071002, China}
\affiliation{Hebei Key Lab of Optic-Electronic Information and Materials, Hebei University, Baoding 071002, China}

\begin{abstract}
We investigate spherically symmetric, steady state, adiabatic accretion onto a Tangherlini-Reissner-Nordstrom black hole in arbitrary dimensions by using $D$-dimensional general relativity. We obtain basic equations for accretion and determine analytically the critical points, the critical fluid velocity, and the critical sound speed. We lay emphasis on the condition under which the accretion is possible. This condition constrains the ratio of mass to charge in a narrow limit, which is independent of dimension for large dimension. This condition may challenge the validity of the cosmic censorship conjecture since a naked singularity is eventually produced as the magnitude of charge increases compared to the mass of black hole.
\end{abstract}

\pacs{04.70.-s, 04.70.Bw, 97.60.Lf}

\maketitle

\section{Introduction}
The extra dimension in physics is especially important, for example, the higher dimensions are needed for consistency in string theory. It believed that the physics of higher dimensional black holes can be markedly different and much richer than in four dimensions. The accretion process could provide possible way to detect the effects of higher dimensions. Accretion of matter onto a black hole is also the most likely scenario to explain the high energy output from active galactic nuclei and quasars, which is an important phenomenon of long-standing interest to astrophysicists. Pressure-free gas dragged onto a massive central object was first considered in \cite{hoyle1939effect,Bondi:1944jm}, which was generalized to the case of the spherical accretion of adiabatic fluids onto astrophysical objects \cite{Bondi:1952ni}. In the framework of general relativity the steady-state spherically symmetric flow of matter into or out of a condensed object was examined by Michel \cite{michel1972accretion}. Since then accretion has been an extensively studied topic in the literature, \citep{Babichev:2004yx, Babichev:2008jb, michel1972accretion, Jamil:2008bc, Rodrigues:2016uor, Babichev:2008dy, JimenezMadrid:2005rk, Bhadra:2011me, Yang:2016sjy, Ganguly:2014cqa, Gao:2008jv, Mach:2013fsa, Mach:2013gia, Karkowski:2012vt, Abbas:2013fha, Abbas:2018ygc, Abbas2019}. Quantum gravity effects of accretion onto a Schwarzschild black hole were considered in the context of asymptotically safe scenario \cite{Yang:2015sfa}. An exact solution was obtained for dust shells collapsing towards a black hole \cite{Liu:2009ts, Zhao:2018ani}. Analytic solution for accretion of a gaseous medium with a adiabatic equation of state ($P=\rho$) was obtained for a moving Schwarzschild black hole and a moving Kerr back hole \cite{petrich1988accretion} and a moving Reissner-Nordstr\"{o}m Black Hole \cite{Jiao:2016uiv}.

Here we will investigate the effects of higher dimension on accretion. An interesting study of higher dimensional accretion onto TeV black holes was discussed in the Newtonian limit \cite{Giddings:2008gr}. The accretion of phantom matter onto 5-dimensional charged black holes was studied in \cite{Sharif:2011ih}. The accretion of phantom energy onto 5-dimensional extreme Einstein-Maxwell-Gauss-Bonnet black hole was investigated in \cite{Jamil:2011sx}. Matter onto a Schwarzschild black hole in arbitrary dimensions was analyzed in \cite{John:2013bqa}. Accretions of dark matter and dark energy onto ($n+2$)-dimensional Schwarzschild black hole and Morris-Thorne wormhole was considered in \cite{Debnath:2015yva}. Steady-state polytropic fluid accretion onto a charged higher-dimensional black hole was dealt in \cite{Sharif:2016pqy}, but the conditions under which the accretion is possible has not yet been discussed. In this paper we will reconsider the problem of matter accreting onto a Tangherlini-Reissner-Nordstrom in arbitrary dimensions, focusing on the conditions under which the accretion will happen. The gravitational model used here is D-dimensional general relativity. We find the conditions give constraints on the ratio of mass to charge.

The paper is organized as follows. In next section, we will present the fundamental equations for matter accreting onto a Tangherlini-Reissner-Nordstrom black hole. In section III, we will consider the critical points and the conditions the critical points must fulfil. Finally, we will briefly summarize and discuss our results.

\section{Basic equations for accretion}
We consider the Tangherlini-Reissner-Nordstrom black hole which is a D-dimensional nonrotating charged black hole and its geometry is represented by the following line element \cite{Tangherlini:1963bw}
\begin{eqnarray}
\label{metric}
ds^2=-g(r)dt^2+g^{-1}(r)dr^2+r^2d\Omega^{2}_{D-2},
\end{eqnarray}
where the mass $M$ is measured at infinity and
\begin{eqnarray}
g(r)=\left[1-\left(\frac{r_{\rm{g}}}{r}\right)^{D-3}+\left(\frac{r_{\rm{Q}}}{r}\right)^{2(D-3)}\right],
\end{eqnarray}
with
\begin{eqnarray}
r_{\rm{g}}=\left[\frac{2kM}{(D-2)\Omega_{D-2}}\right]^{\frac{1}{(D-3)}},\\
r_{\rm{Q}}=\left[\frac{kQ^{2}}{(D-2)(D-3)\Omega^{2}_{D-2}}\right]^\frac{1}{2(D-3)},
\end{eqnarray}
where $k=8\pi G$. For convenience, we introduce the gravitational radius $r_{\rm{g}}$ and the charge radius $r_{\rm{Q}}$ in D-dimensional spacetime. We take the units $c=G=1$ and use the comoving coordinates $x^\mu=(t,r,\theta_1,\theta_2,...,\theta_{D-2})$. $\Omega_{D-2}$ is an unit (D-2)-dimensional sphere
\begin{eqnarray}
\Omega_{D-2}=\frac{2\pi^{\frac{D-1}{2}}}{\Gamma\big(\frac{D-1}{2}\big)},\nonumber
\end{eqnarray}
and $d\Omega^{2}_{D-2}$ is the line element on the sphere
\begin{eqnarray}
d\Omega^{2}_{D-2}=d\theta^{2}_{1}+\sum^{D-2}_{n=2}\left(\prod^{n}_{m=2}sin^{2}\theta_{m-1}\right)d\theta^{2}_{n}\nonumber.
\end{eqnarray}
The above metric (\ref{metric}) describes a static, unrotating, higher dimensional charged black holes. If $r_{\rm{g}}^{2(D-3)}<4r_{\rm{Q}}^{2(D-3)}$, the metric (\ref{metric}) has the naked singularity at $r=0$, this situation is similar to the Reissner-Nordstrom solution in the 4-dimensional spacetime when $r_{\rm{g}}<2r_{\rm{Q}}$. If $r_{\rm{g}}^{2(D-3)}\geq 4r_{\rm{Q}}^{2(D-3)}$, the metric (\ref{metric}) at
\begin{eqnarray}
r_{\rm{H}\pm}\equiv\left[\frac{1}{2}\left(r_{\rm{g}}^{D-3}\pm\sqrt{r_{\rm{g}}^{2(D-3)}-4r_{\rm{Q}}^{2(D-3)}}\right)\right]^{1/(D-3)}\nonumber
\end{eqnarray}
has black hole horizons, this situation is also similar to the Reissner-Nordstrom black hole if $r_{g}\geq2r_{Q}$.

We consider a static radial matter flow onto the Tangherlini-Reissner-Nordstrom black hole with mass $M$. the matter is approximated as an ideal flow described by the following energy-momentum tensor
\begin{eqnarray}
T^{\alpha\beta}=(\rho+p)u^{\alpha}u^{\beta}+pg^{\alpha\beta},
\end{eqnarray}
where $\rho$ is the fluid proper energy density and $p$ is the fluid proper pressure. The D-velocity of the fluid
$u^{\alpha}=dx^{\alpha}/ds$ obeys the normalization condition $u^{\alpha}u_{\alpha}=-1$.
In the meantime, we define the proper baryon numbers density $n$ and then baryon numbers flow $J^{\alpha}=nu^{\alpha}$.
Ignoring the self-gravity of the flow, all of these quantities are measured in the flow's local inertial frame. Assuming no particles are generated or destroyed, meaning the number of particles is conserved, we have
\begin{eqnarray}
\label{particle}
J^{\alpha}_{~~;\alpha}=(nu^{\alpha})_{;\alpha}=0,
\end{eqnarray}
where $;a$ denotes the covariant derivative with respect to the coordinate $x^\alpha$. The conservation of energy and momentum is determined by
\begin{eqnarray}
\label{enm}
T^{\alpha\beta}_{~~~;\alpha}=0,
\end{eqnarray}
We denote the radial component of the $D$-velocity as $v(r)=u^{1}=dr/ds$. Since the normalization condition for $u^\alpha$ and
the velocity component is zero for $\alpha>1$, one has
\begin{eqnarray}
(u^{0})^{2}=\frac{\upsilon^{2}+1-\left(\frac{r_{\rm{g}}}{r}\right)^{D-3}+\left(\frac{r_{\rm{Q}}}{r}\right)^{2(D-3)}}{\left[1-\left(\frac{r_{\rm{g}}}{r}\right)^{D-3}+
\left(\frac{r_{\rm{Q}}}{r}\right)^{2(D-3)}\right]^{2}},
\end{eqnarray}
In $D$-dimensional Tangherlini-Reissner-Nordstrom black hole, the equation (\ref{particle}) can be rewritten as
\begin{eqnarray}
\label{particle1}
\frac{1}{r^{D-2}}\frac{d}{dr}\left(r^{D-2}n\upsilon\right)=0,
\end{eqnarray}
For spherical symmetry and steady-state flow, the $\beta=0$ component of equation (\ref{enm}) gives
\begin{eqnarray}
\label{comp0}
\frac{1}{r^{D-2}}\frac{d}{dr}\left[r^{D-2}\left(\rho+p\right)\upsilon\sqrt{1-\left(\frac{r_{\rm{g}}}{r}\right)^{D-3}+\left(\frac{r_{\rm{Q}}}{r}\right)^{2(D-2)}+\upsilon^{2}}\right]=0.
\end{eqnarray}
After some calculations, the $\beta=1$ component of equation (\ref{enm}) can be simplified as
\begin{eqnarray}
\label{1comp}
\upsilon\frac{d\upsilon}{dr}=-\frac{dp}{dr}
\frac{1-\left(\frac{r_{\rm{\rm{g}}}}{r}\right)^{D-3}+\left(\frac{r_{\rm{Q}}}{r}\right)^{2(D-3)}+\upsilon^{2}}{\rho+p}
-\frac{D-3}{2}\left(\frac{r_{\rm{g}}^{D-3}}{r^{D-2}}-\frac{2r^{2(D-3)}_{\rm{Q}}}{r^{2D-5}}\right),
\end{eqnarray}
These expressions generalize, to arbitrary dimensions $D$, the results acquired in \cite{Jamil:2008bc} for spherical accretion onto a Reissner-Nordstrom black hole and can reduce to the results in 4-dimension
\begin{eqnarray}
\frac{1}{r^{2}}
\frac{d}{dr}\left(r^{2}n\upsilon\right)=0,
\end{eqnarray}
\begin{eqnarray}
\frac{1}{r^{2}}
\frac{d}{dr}\left[r^{2}(\rho+p)\upsilon\left(1-\frac{2M}{r}+\frac{Q^{2}}{r^{2}}+\upsilon^{2}\right)^{1/2}\right]=0,
\end{eqnarray}
\begin{eqnarray}
\upsilon\frac{d\upsilon}{dr}=-\frac{dp}{dr}\left(\frac{1-\frac{2M}{r}+\frac{Q^{2}}{r^{2}}+\upsilon^{2}}{\rho+p}\right)
-\left(\frac{M}{r^{2}}-\frac{Q^{4}}{r^{3}}\right).
\end{eqnarray}
Writing the continuity equation explicitly in the form of a conservation equation, integrating equation (\ref{particle1}) over $(D-1)$-dimensional volume, and multiplying it with $m$ which denotes the mass of each gas particle, we have
\begin{eqnarray}
\label{mass}
\dot{M}=\frac{2\pi^{(D-1)/2}}{\Gamma(\frac{D-1}{2})}r^{D-2}mn\upsilon,
\end{eqnarray}
where $\dot{M}$ is an integration constant which has dimensions of mass per unit time, meaning it is the higher dimensional generalization of mass accretion rate. Combining equations (\ref{particle1}) and (\ref{comp0}) yields:
\begin{eqnarray}
\label{comb}
\left(\frac{\rho+p}{n}\right)^{2}\left[1-\left(\frac{r_{\rm{g}}}{r_{\rm{c}}}\right)^{D-3}+\left(\frac{r_{\rm{Q}}}{r_{\rm{c}}}\right)^{2(D-3)}+\upsilon^{2}\right]=
\left(\frac{\rho_{\infty}+p_{\infty}}{n_{\infty}}\right)^{2},
\end{eqnarray}
where the subscript ``$\infty$" denotes the corresponding asymptotic values at infinity, these values are usually used as the initial values of the fluid. The equations (\ref{mass}) and (\ref{comb}) are the basic conservation equations for the material flow onto a $D$-dimensional
Tangherlini-Reissner-Nordstrom black hole where the back-reaction of matter is ignored.

\section{Analysis for the accretion}
From the qualitative analysis of equations (\ref{particle1}) and (\ref{1comp}), we can calculate the rate of mass accretion. For the adiabatic flow, there is no entropy generation, the mass-energy conservation is given by
\begin{eqnarray}
\label{conser}
0=Tds=d \left(\frac{\rho}{n}\right)+p d\left(\frac{1}{n}\right),
\end{eqnarray}
from which one can easily get the following relation
\begin{eqnarray}
\label{men}
\frac{d\rho}{dn}=\frac{\rho+p}{n},
\end{eqnarray}
Like in the four-dimension case, we can define the adiabatic sound speed $a$ as
\begin{eqnarray}
\label{ss}
a^{2}\equiv\frac{dp}{d\rho}=\frac{dp}{dn}
\frac{n}{\rho+p},
\end{eqnarray}
where equation (\ref{men}) has been used. For Tangherlini-Reissner-Nordstrom black hole, the continuity equation (\ref{particle1}) and the momentum equation (\ref{1comp}) can be rewritten as (similar calculating process see \cite{Yang:2016sjy})
\begin{eqnarray}
\frac{1}{\upsilon}\upsilon^{\prime}+\frac{1}{n}n^{\prime}=-\frac{D-2}{r},
\end{eqnarray}
\begin{eqnarray}
\upsilon\upsilon^{\prime}+\left[1-\left(\frac{r_{\rm{g}}}{r}\right)^{D-3}+\left(\frac{r_{\rm{Q}}}{r}\right)^{2(D-3)}+\upsilon^{2}\right]\frac{a^{2}}{n}n^{\prime}=
-\frac{D-3}{2}\left[\frac{r_{\rm{g}}^{D-3}}{r^{D-2}}-\frac{2r_{\rm{Q}}^{2(D-3)}}{r^{2D-5}}\right],
\end{eqnarray}
where the prime $(\prime)$ represents a derivative with respect to $r$.  From these equations we get the following system
\begin{eqnarray}
\label{N1}
\upsilon^{\prime}&=&\frac{N_{1}}{N},\\
\label{N2}
n^{\prime}&=&-\frac{N_{2}}{N},
\end{eqnarray}
where
\begin{eqnarray}
N_{1}=\frac{1}{n}
\left\{\left[1-\left(\frac{r_{\rm{g}}}{r}\right)^{D-3}+\left(\frac{r_{\rm{Q}}}{r}\right)^{2(D-3)}+\upsilon^{2}\right]
\left(D-2\right)\frac{a^{2}}{r}-\frac{D-3}{2}\left[\frac{r_{\rm{g}}^{D-3}}{r^{D-2}}-\frac{2r_{\rm{Q}}^{2(D-3)}}{r^{2D-5}}\right]\right\},
\end{eqnarray}
\begin{eqnarray}
N_{2}=\frac{1}{\upsilon}\left\{(D-2)
\frac{\upsilon^{2}}{r}-\frac{D-3}{2}\left[\frac{r_{\rm{g}}^{D-3}}{r^{D-2}}-\frac{2r_{\rm{Q}}^{2(D-3)}}{r^{2D-5}}\right]\right\},
\end{eqnarray}
\begin{eqnarray}
\label{den}
N=\frac{\upsilon^{2}-\left[1-\left(\frac{r_{\rm{g}}}{r}\right)^{D-3}+\left(\frac{r_{\rm{Q}}}{r}\right)^{2(D-3)}+\upsilon^{2}\right]a^{2}}{\upsilon
n},
\end{eqnarray}
For $r\rightarrow +\infty$, we require that the fluid is subsonic, that is $v<a$ and the speed of sound is less than the speed of light, that is $a<1$, we have $v^{2}\ll1$  the denominator (\ref{den}) can be simplified as
\begin{eqnarray}
N\approx\frac{\upsilon^{2}-a^{2}}{\upsilon
n},
\end{eqnarray}
meaning $N>0$ for large $r$. At the outer horizon,
$1-(r_{\rm{g}}/r)^{D-3}+(r_{\rm{Q}}/r)^{2(D-3)}=0$, one has
\begin{eqnarray}
N=\frac{\upsilon^{2}(1-a^{2})}{\upsilon
n},
\end{eqnarray}
Thinking of the causality constraint $a^{2}<1$, we have $N<0$, therefore there must exist critical points $r_{\rm{c}}$ at which $N=0$.
The flow must pass through each critical point of spacetime, but at these points the denominators on the right hand sides of the equations (\ref{N1}) and (\ref{N2}) are zero, their numerators $N_1$ and $N_2$ must also be zero to avoid discontinuities in the flow. These equations determine the critical point
\begin{eqnarray}
N_{1}=\frac{1}{n_{\rm{c}}}
\left\{\left[1-\left(\frac{r_{\rm{g}}}{r_{\rm{c}}}\right)^{D-3}+\left(\frac{r_{\rm{Q}}}{r_{\rm{c}}}\right)^{2(D-3)}+\upsilon^{2}_{\rm{c}}\right]
(D-2)\frac{a^{2}_{\rm{c}}}{r_{\rm{c}}}-\frac{D-3}{2}\left[\frac{r_{\rm{g}}^{D-3}}{r^{D-2}_{\rm{c}}}-\frac{2r_{\rm{Q}}^{2(D-3)}}{r^{2D-5}_{\rm{c}}}\right]\right\}=0,
\end{eqnarray}
\begin{eqnarray}
\label{n11}
N_{2}=\frac{1}{\upsilon_{\rm{c}}}\left[(D-2)
\frac{\upsilon^{2}_{\rm{c}}}{r_{\rm{c}}}-\frac{D-3}{2}\left(\frac{r_{\rm{g}}^{D-3}}{r^{D-2}_{\rm{c}}}-\frac{2r_{\rm{Q}}^{2(D-3)}}{r^{2D-5}_{\rm{c}}}\right)\right]=0,
\end{eqnarray}
\begin{eqnarray}
\label{n22}
N=\frac{\upsilon^{2}_{\rm{c}}-\left[1-\left(\frac{r_{\rm{g}}}{r_{\rm{c}}}\right)^{D-3}+\left(\frac{r_{\rm{Q}}}{r_{\rm{c}}}\right)^{2(D-3)}+\upsilon^{2}_{\rm{c}}\right]
a^{2}_{\rm{c}}}{\upsilon_{\rm{c}}
n_{\rm{c}}}=0,
\end{eqnarray}
where $a_{\rm{c}}\equiv a(r_{\rm{c}}), v_{\rm{c}}\equiv v(r_{\rm{c}})$, etc. Through the equations (\ref{n11}) and (\ref{n22}), we get the critical radial velocity, the critical sound speed, and the critical radius, respectively, as
\begin{eqnarray}
\label{cv}
\upsilon^{2}_{\rm{c}}=\frac{1}{2}
\frac{D-3}{D-2}\left[\left(\frac{r_{\rm{g}}}{r_{\rm{c}}}\right)^{D-3}-2\left(\frac{r_{\rm{Q}}}{r_{\rm{c}}}\right)^{2(D-3)}\right],
\end{eqnarray}
\begin{eqnarray}
\label{cs}
a^{2}_{\rm{c}}=\frac{\upsilon^{2}_{\rm{c}}}{1-\left(\frac{r_{\rm{g}}}{r_{\rm{c}}}\right)^{D-3}+\left(\frac{r_{\rm{Q}}}{r_{\rm{c}}}\right)^{2(D-3)}+\upsilon^{2}_{\rm{c}}},
\end{eqnarray}
and
\begin{eqnarray}
r^{D-3}_{\rm{c}}=\frac
{[(D-1)a^{2}_{\rm{c}}+D-3]r^{D-3}_{g}\pm\sqrt{[(D-1)a^{2}_{\rm{c}}+D-3]^{2}r^{2(D-3)}_{\rm{g}}-16(D-2)a^{2}_{\rm{c}}(a^{2}_{\rm{c}}+D-3)r^{2(D-3)}_{\rm{Q}}}}
{4(D-2)a^{2}_{\rm{c}}}.
\end{eqnarray}
where ``$+$" corresponding the critical points lying out of the event horizon, while ``$-$" corresponding the critical points lying in the black hole. The solutions of equations (\ref{cv}) and (\ref{cs}) must satisfy the conditions $v^{2}_{\rm{c}}\geq0$ and $a^{2}_{\rm{c}}\geq0$, that is
\begin{eqnarray}
\left\{
\begin{aligned}
\upsilon^{2}_{\rm{c}}&=\frac{1}{2}
\frac{D-3}{D-2}\left[\left(\frac{r_{\rm{g}}}{r_{\rm{c}}}\right)^{D-3}-2\left(\frac{r_{\rm{Q}}}{r_{\rm{c}}}\right)^{2(D-3)}\right]\geq0,\\
a^{2}_{\rm{c}}&=\frac{\upsilon^{2}_{\rm{c}}}{1-\left(\frac{r_{\rm{g}}}{r_{\rm{c}}}\right)^{D-3}+\left(\frac{r_{\rm{Q}}}{r_{\rm{c}}}\right)^{2(D-3)}+\upsilon^{2}_{\rm{c}}}
=\frac{(D-3)\left[\left(\frac{r_{\rm{g}}}{r_{\rm{c}}}\right)^{D-3}-2\left(\frac{r_{\rm{Q}}}{r_{\rm{c}}}\right)^{2(D-3)}\right]}
{2(D-2)-(D-1)\left(\frac{r_{\rm{g}}}{r_{\rm{c}}}\right)^{D-3}+2\left(\frac{r_{\rm{Q}}}{r_{\rm{c}}}\right)^{2(D-3)}} \geq 0,\\
\end{aligned}
\right.
\end{eqnarray}
which can be reduced to
\begin{eqnarray}
\label{inq1}
r_{\rm{g}}^{D-3}r_{\rm{c}}^{D-3}-2r_{\rm{Q}}^{2(D-3)}\geq0,
\end{eqnarray}
and
\begin{eqnarray}
\label{inq2}
2(D-2)r_{\rm{c}}^{2(D-3)}-(D-1)r_{\rm{g}}^{D-3}r_{\rm{c}}^{D-3}+2r_{\rm{Q}}^{2(D-3)}\geq0.
\end{eqnarray}
Inequality (\ref{inq2}) can be factorized as
\begin{eqnarray}
2(D-2)r_{\rm{c}}^{2(D-3)}-(D-1)r_{\rm{g}}^{D-3}r_{\rm{c}}^{D-3}+2r_{\rm{Q}}^{2(D-3)}=2(D-2)(r_{\rm{c}}^{D-3}-r_{\rm{c+}}^{D-3})(r_{\rm{c}}^{D-3}-r_{\rm{c-}}^{D-3})\geq0,
\end{eqnarray}
where
\begin{eqnarray}
r_{\rm{c}\pm}^{D-3}=\frac{(D-1)r_{\rm{g}}^{D-3}\pm\sqrt{(D-1)^{2}r_{\rm{g}}^{2(D-3)}-16(D-2)r_{\rm{Q}}^{2(D-3)}}}{4(D-2)}.
\end{eqnarray}
It is obvious that these two roots will be real value if $(D-1)r_{\rm{g}}^{2(D-3)}-16(D-2)r_{\rm{Q}}^{2(D-3)}\geq0$ which can also be written as
\begin{eqnarray}
\label{inqr1}
\frac{r_{\rm{g}}^{2(D-3)}}{r_{\rm{Q}}^{2(D-3)}}\geq \frac{16(D-2)}{(D-1)^{2}}.
\end{eqnarray}
Therefore $r_{\rm{c}\pm}^{D-3}$ are positive and satisfy $r_{\rm{c+}}^{D-3}>r_{\rm{c-}}^{D-3}>0$. In general, for $4r_{\rm{Q}}^{2(D-3)}\leq r_{\rm{g}}^{2(D-3)}$, the inner critical points will lie between $r_{\rm{H-}}^{D-3}\leq r_{\rm{c}}^{D-3}\leq r_{\rm{H+}}^{D-3}$, while the outer critical points will satisfy $r_{\rm{c}}^{D-3}\geq r_{\rm{H}+}^{D-3}$. From (\ref{inq2}),we can get the critical points are located in two regions: (1) $r_{\rm{c}}^{D-3} > r_{\rm{c+}}^{D-3}$ or (2) $ 0 < r_{\rm{c}}^{D-3} < r_{\rm{c-}}^{D-3}$.
In order to get the solution of the critical point, we substitute $r_{\rm{c\pm}}^{D-3}$ in (\ref{inq1}). First we consider case (1) $r_{\rm{c}}^{D-3} > r_{\rm{c+}}^{D-3}$. Inserting $r_{\rm{c+}}^{D-3}$, inequality (\ref{inq1}) gives
\begin{eqnarray}
\label{inq3}
r_{\rm{g}}^{D-3}\sqrt{(D-1)r_{\rm{g}}^{2(D-3)}-16(D-2)r_{\rm{Q}}^{2(D-3)}}\geq 8(D-2)r_{\rm{Q}}^{2(D-3)}-(D-1)r_{\rm{g}}^{2(D-3)}.
\end{eqnarray}
Since $r_{\rm{g}}^{D-3}>0$, if $8(D-2)r_{\rm{Q}}^{2(D-3)}-(D-1)r_{\rm{g}}^{2(D-3)}<0$, inequality (\ref{inq3}) does not yield any physical solutions. So we must have
$8(D-2)r_{\rm{Q}}^{2(D-3)}-(D-1)r_{\rm{g}}^{2(D-3)}>0$, which yields
\begin{eqnarray}
\label{inqr2}
\frac{r_{\rm{g}}^{2(D-3)}}{r_{\rm{Q}}^{2(D-3)}}<\frac{8(D-2)}{D-1}.
\end{eqnarray}
Squared (\ref{inq3}) on both sides we obtain
\begin{eqnarray}
\label{inqr3}
\frac{r_{\rm{g}}^{2(D-3)}}{r_{\rm{Q}}^{2(D-3)}}\geq4.
\end{eqnarray}
Combining the inequalities (\ref{inqr1}), (\ref{inqr2}) and (\ref{inqr3}) implies
\begin{eqnarray}
\label{res}
4\leq\frac{r_{\rm{g}}^{2(D-3)}}{r_{\rm{Q}}^{2(D-3)}}<\frac{8(D-2)}{D-1}.
\end{eqnarray}
Thus, the accretion is allowed through the critical point $r_{\rm{c+}}^{D-3}$ if the inequality (\ref{res}) is satisfied.
Now we consider case (2), $ 0 < r_{\rm{c}}^{D-3} < r_{\rm{c-}}^{D-3}$. Substituting $r_{\rm{c-}}^{D-3}$ into (\ref{inq1}) gives
\begin{eqnarray}
r_{\rm{g}}^{D-3}\sqrt{(D-1)r_{\rm{g}}^{2(D-3)}-16(D-2)r_{\rm{Q}}^{2(D-3)}}\leq (D-1)r_{\rm{g}}^{2(D-3)}-8(D-2)r_{\rm{Q}}^{2(D-3)}.
\end{eqnarray}
which is satisfied if
\begin{eqnarray}
\label{2inq}
\frac{r_{\rm{g}}^{2(D-3)}}{r_{\rm{Q}}^{2(D-3)}}>\frac{8(D-2)}{D-1},
\end{eqnarray}
and
\begin{eqnarray}
\label{3inq}
\frac{r_{\rm{g}}^{2(D-3)}}{r_{\rm{Q}}^{2(D-3)}} \leq4.
\end{eqnarray}
The inequalities (\ref{2inq}) and (\ref{3inq}) are mutually contradictory, meaning that there is no solution for $r_{\rm{c}}$ in case (2). So accretion is not possible through $r_{\rm{c-}}$. This critical point yields the ratio of mass to charge of the black hole in the range specified by (\ref{res}), which allows accretion onto the charged spherically symmetric higher dimensional black holes. For $D\rightarrow +\infty$, the inequality (\ref{res}) reduces to $4\leq\frac{r_{\rm{g}}^{2(D-3)}}{r_{\rm{Q}}^{2(D-3)}}<8$, which is independent of the dimension. Since a naked singularity is eventually produced as magnitude of charge increases compared to mass of black hole, the inequality (\ref{res}) may also challenge the validity of the cosmic censorship conjecture. We note here that using $Q=0$ in the inequality (\ref{res}) to derive the condition for the Tangherlini black hole ($D$-dimensional Schwarzschild black hole) can be misleading, one has $r_{\rm{Q}}=0$ in this case and the inequality (\ref{res}) can not held any longer, since it is obtained by using the outer apparent horizon and critical points, while Tangherlini black hole possesses unique horizon and the critical point. For $D=4$, however, the inequality (\ref{res}) reduce to the results obtained in \cite{Jamil:2008bc}. A similar constrain appears in the discussion of pseudo-Newtonian forces \cite{Qadir1986The}.

\section{Conclusions and discussions}
In this paper we formulated and solved the problem of spherically symmetric, steady state, adiabatic accretion onto a Tangherlini-Reissner-Nordstrom black hole in arbitrary dimensions by using $D$-dimensional general relativity. We obtained basic equations for accretion and determine analytically the critical points, the critical fluid velocity, the critical sound speed, and subsequently the mass accretion rate. We found that accretion is possible only through $r_{\rm{c+}}$ which yields a limit on the ratio of mass to charge given by the inequality (\ref{res}). This inequality incorporates both extremal and non-extremal higher dimension charged black holes. It predicts the existence of large charges onto black holes, although no such evidence has been successfully deduced from the astrophysical observations, but it is consistently deduced by the $D$-dimensional general relativity, as shown here. The inequality (\ref{res}) is independent of dimension for lager dimension. It may challenge the validity of the cosmic censorship conjecture since a naked singularity is eventually produced as magnitude of charge increases compared to mass of black hole. It is worth investigating whether this analysis can be extended for a Kerr-Neumann black hole.

\begin{acknowledgments}
This study is supported in part by National Natural Science Foundation of China (Grant Nos. 11273010), Hebei Provincial Natural Science Foundation of China (Grant No. A2014201068), the Outstanding Youth Fund of Hebei University (No. 2012JQ02), and the Midwest universities comprehensive strength promotion project.
\end{acknowledgments}

\bibliographystyle{ieeetr}
\bibliography{ref}

\end{document}